\documentclass[aps,pra,superscriptaddress,showpacs,twocolumn]{revtex4-2}
\usepackage{bm}
\usepackage{amsmath}
\usepackage{float}
\allowdisplaybreaks
\usepackage{longtable}
\usepackage{dcolumn}
\newcolumntype{.}{D{x}{}{-1}}

\newcommand*{\centt}[1]{\multicolumn{1}{c}{#1}}
\newcolumntype{w}[1]{D{.}{.}{#1}}

\usepackage{graphicx}
\usepackage{xcolor}
\def\mynote1#1{{\color{blue}{\textsc{\footnotesize Question/Comment:} #1}}}
\def\mynote2#1{{\color{magenta}{#1}}}

\newcommand{\lbr}{\langle}
\newcommand{\rbr}{\rangle}

\usepackage{nicefrac}

\begin{document}
\title{Complete $\bm{\alpha^7\,m}$ Lamb shift of helium triplet states}

\author{Vojt\v{e}ch Patk\'o\v{s}}
\affiliation{Faculty of Mathematics and Physics, Charles University,  Ke Karlovu 3, 121 16 Prague 2, Czech Republic}

\author{Vladimir A. Yerokhin}
\affiliation{Center for Advanced Studies, Peter the Great St.~Petersburg Polytechnic University,
Polytekhnicheskaya 29, 195251 St.~Petersburg, Russia}

\author{Krzysztof Pachucki}
\affiliation{Faculty of Physics, University of Warsaw,
             Pasteura 5, 02-093 Warsaw, Poland}

\date{\today}

\begin{abstract}
We have derived the complete formula for the $\alpha^7\,m$ contribution to energy levels of an
arbitrary triplet state of the helium atom, performed numerical calculations for the $2^3S$ and
$2^3P$ states, and thus improved the theoretical accuracy of ionization energies of these
states by more than an order of magnitude. Using the nuclear charge radius extracted from the
muonic helium Lamb shift, we obtain the theoretical prediction in excellent agreement with the
measured $2^3S - 2^3P$ transition energy [X.~Zheng et al., Phys. Rev. Lett. {\bf 199}, 263002
(2017)]. At the same time we observe significant discrepancies with experiments for the $2^3S-
3^3D$ and $2^3P - 3^3D$ transitions.
\end{abstract}
\maketitle

\section{Introduction}
High precision spectroscopic measurements in atoms and molecules can be used for the determination of
fundamental constants such as the Rydberg constant \cite{mohr:20:codata}  and the electron-nuclear mass ratio \cite{HDion1, HDion2}. 
They can also be used for the determination of nuclear properties, among them magnetic dipole
and electric quadrupole moments. In the present work we investigate the possibility of
determining the nuclear charge radius by means of atomic spectroscopy.

The differences of the (squares of) nuclear charge radii between different isotopes are routinely
determined nowadays from measurements of the isotope shifts of transition frequencies
\cite{lu:13,maass:19,pachucki:15:jpcrd, blaum}. Here we address a more ambitious task of determining the
absolute value of the nuclear charge radius, specifically, that of the helium atom. The main
motivation of the spectroscopic determination of nuclear radii is to make possible a comparison
of different methods, such as electron scattering and the muonic-atom spectroscopy, and to search
for possible deviations that might signal the existence of unknown interactions at the atomic
scale.

The spectroscopic determination of the nuclear radius has already been accomplished for the
hydrogen atom. Importantly, it was performed by two independent methods: from ordinary hydrogen
\cite{beyer:17,fleurbaey:18, bezginov:19, grinin:20} and from muonic hydrogen \cite{pohl:10,antognini:13}. At first the
comparison of the two methods revealed a large discrepancy, which became known as the proton size
puzzle. This discrepancy seems to be close to a resolution now \cite{karr:20} because several
recent spectroscopic and scattering experiments showed to be consistent with the muonic hydrogen
proton radius. As a result, the comparison of ordinary and muonic hydrogen has provided improved
values for the proton radius and the Rydberg constant and forced a reconsideration of systematic
effects in hydrogen spectroscopy.

One may expect that a similar comparison performed for other nuclei will also reveal interesting
findings. An important step towards such a comparison is the recent muonic helium experiment
\cite{krauth:21}, which determined the charge radius of the helium-4 nucleus (the $\alpha$
particle) with a 0.05\% precision.

The goal of the present work is to improve the theoretical accuracy of the $2^3S-2^3P$ transition
energy in atomic helium to a level sufficient for the determination of the nuclear charge radius
from the existing measurements in ordinary helium. We achieve this by performing the complete
calculation of the $\alpha^7\,m$ QED effects. Unfortunately, we also find that our calculation
does not resolve the previously reported discrepancy of theoretical predictions with experimental
results for the $2^3S-2^3D$ and $2^3P-2^3D$ transitions \cite{wienczek:19}. In view of this, we
postpone the determination of the $\alpha$-particle charge radius until these discrepancies are resolved.
Henceforth, we present our calculations of the complete $\alpha^7\,m$ QED effects and obtain the improved
theoretical predictions for atomic helium energy levels using the quantum electrodynamic theory.

\section{Perturbative expansion of atomic energy levels}
The basic assumption in bound-state quantum electrodynamics is the possibility of the  expansion of the bound
state energy $E$ in a power series of the fine-structure constant $\alpha$,
\begin{eqnarray}
E\Bigl(\alpha, \frac{m}{M}\Bigr) &=& \alpha^2\,E^{(2)}\Bigl(\frac{m}{M}\Bigr)
+ \alpha^4\,E^{(4)}\Bigl(\frac{m}{M}\Bigr)
+ \alpha^5\,E^{(5)}\Bigl(\frac{m}{M}\Bigr)
 \nonumber \\&&
+ \alpha^6\,E^{(6)}\Bigl(\frac{m}{M}\Bigr)
+ \alpha^7\,E^{(7)}\Bigl(\frac{m}{M}\Bigr) +\ldots \,,\label{04}
\end{eqnarray}
where $m/M$  is the electron-to-nucleus mass ratio and the expansion coefficients $E^{(n)}$ may
contain finite powers of $\ln\alpha$. The coefficients $E^{(i)}(m/M)$ are further expanded in
powers of the $m/M$ ratio,
\begin{equation}
E^{(i)}\Bigl(\frac{m}{M}\Bigr) = E^{(i,0)} + \frac{m}{M}\,E^{(i,1)} + \Bigl(\frac{m}{M}\Bigr)^2\,E^{(i,2)} + \ldots\,. \label{05}
\end{equation}

The leading expansion term $E_0 \equiv E^{(2,0)}$ is the nonrelativistic energy, which is the
eigenvalue of the nonrelativistic Hamiltonian $H_0$. For the helium atom,
\begin{equation}
H_0 = \frac{p_1^2}{2}+ \frac{p_2^2}{2} - \frac{Z}{r_1} - \frac{Z}{r_2} + \frac{1}{r}\,,
\end{equation}
where $r = |\vec{r}_1-\vec{r}_2|$. Further expansion terms in Eqs.~(\ref{04}) and (\ref{05}) can
be expressed as expectation values of some effective Hamiltonians with the nonrelativistic wave
function. The derivation of the effective Hamiltonians is the central problem, and this can be
accomplished within the approach of the nonrelativistic QED (NRQED), which is employed here.
While the leading-order terms are simple, the derivation becomes increasingly complicated
for high powers of $\alpha$. The complete theory of helium energy levels up to order
$\alpha^6\,m$ was reviewed in our former work \cite{pachucki:17:heSummary}. In the present work,
we summarize the $\alpha^7\,m$ contribution and perform its numerical calculations for the $2^3S$
and $2^3P$ states. In this calculation we assume the infinitely heavy nucleus. The corresponding
finite nuclear mass corrections are much smaller than the uncertainty due to the approximate calculation
of the next order $\alpha^8\,m$ contribution, and therefore they are neglected.

\section{$\bm{\alpha^7\,m}$ contribution}

The $\alpha^7m$ contribution $E^{(7)}$ is represented as a sum of three terms,
\begin{align}\label{eq:0}
  E^{(7)} =   E_L^{(7)} + E^{(7)}_{\rm exch} + E^{(7)}_{\rm rad}\,,
\end{align}
where $E_L^{(7)}$ is the low-energy part~-~specifically, the relativistic correction to the
so-called Bethe logarithm; $E^{(7)}_{\rm exch}$ is the part induced by the electron-electron and
electron-nucleus photon exchange; and $E^{(7)}_{\rm rad}$ is induced by the radiative QED effects
beyond those accounted for by $E_L^{(7)}$. Both $E^{(7)}_{\rm exch}$ and $E^{(7)}_{\rm rad}$ have
the same general structure, being the sum of the first-order and second-order perturbation
corrections,
\begin{equation}
E^{(7)}_{\rm exch/rad} = \langle H^{(7)}_{\rm exch/rad} \rangle + 2\,\biggl\langle H^{(4)}\,\frac{1}{(E_0 - H_0)'}\,H^{(5)}_{\rm exch/rad}\biggr\rangle\,.
\end{equation}
Here,  $H^{(4)}$ is the leading relativistic Breit Hamiltonian (see Eq. (7) of Ref. \cite{patkos:20})
and $H^{(5)}$ is the QED $\alpha^5m$ Hamiltonian.

The relativistic correction to the Bethe logarithm was derived and calculated numerically in
Ref.~\cite{yerokhin:18:betherel}, the photon-exchange contribution was derived in
Ref.~\cite{patkos:20}, and the radiative contribution was recently derived in
Ref.~\cite{patkos:21}.

\subsection{Relativistic correction to the Bethe logarithm} We start with the low-energy part in the
leading QED contribution. The leading nonrelativistic (dipole) low-energy contribution of order
$\alpha^5\,m$ is given by
\begin{eqnarray}
E_{L0}(\Lambda) &=& e^2\int_{k<\Lambda} \frac{d^3 k}{(2\,\pi)^3\,2\,k}\,
\left(\delta^{ij}-\frac{k^i\,k^j}{k^2}\right)\,
\nonumber \\ &&\times
\left< P^i\,\frac{1}{E_0-H_0-k}\,P^j \right>\,,
\label{EL0}
\end{eqnarray}
where $\vec P = \vec p_1 + \vec p_2$ and $\Lambda = \lambda\,\alpha^2$ is the high-momentum
cutoff. $E_{L0}(\Lambda)$ diverges when $\lambda \to \infty$, due to the presence of terms
proportional to $\lambda$ and $\ln\lambda$. We obtain the finite part of  $E_{L0}$ by subtracting
all these $\lambda$ dependent terms. The result is by definition the low-energy $m\alpha^5$
contribution, also known as the Bethe logarithm.

The relativistic correction to the Bethe logarithm, $E_L^{(7)}$, is obtained similarly. It
consists of three parts,
\begin{equation}
  E_L^{(7)} = E_{L1} + E_{L2} + E_{L3}\,.
\end{equation}
The first part $E_{L1}$ is a perturbation of the nonrelativistic low-energy contribution $E_{L0}$
in Eq.~(\ref{EL0}) by the Breit Hamiltonian $H^{(4)}$, the second part $E_{L2}$ is induced by the
relativistic correction to the current operator $\vec P/m$, and the third term $E_{L3}$ is the
retardation correction. All of these corrections are defined as  remainders after dropping
$\lambda$-divergent terms $\sim\lambda^2,\lambda,\ln\lambda$, and $\ln^2\lambda$. The divergent
terms are cancelled when combined with corresponding terms from the other contributions in
Eq.~(\ref{eq:0}).

Numerical results for $E_{L1}$, $E_{L2}$ and $E_{L3}$ are taken from
Ref.~\cite{yerokhin:18:betherel} and are summarized in Table \ref{tab:1}. Numerical uncertainties
are negligible in comparison to uncertainties due to higher order corrections.

\begin{table}
\caption{Relativistic corrections to the Bethe logarithm for the $2^3\!S$ and
$2^3\!P$ (centroid) states of helium, in units of $\alpha^7\,m$. \label{tab:1}
}
\begin{ruledtabular}
\begin{tabular}{l..}
\multicolumn{1}{l}{Term} & \multicolumn{1}{c}{$2^3S$} & \multicolumn{1}{c}{$2^3P$} \\
\hline\\[-5pt]
$E_{L1}$ &   -45.129x1\,(35)            &    -41.717x5\,(40)           \\[5pt]
$E_{L2}$  &   335.867x5\,(36)           &  319.160x1\,(36)            \\[5pt]
$E_{L3}$  &  -1\,095.043x\,9\,(3)       &  -1\,045.271x\,(8)         \\
\end{tabular}
\end{ruledtabular}
\end{table}

\subsection{Photon-exchange contribution}

The contribution $E^{(7)}_{\rm exch}$ is induced by the electron-electron and
electron-nucleus photon exchanges, {\em i.e.}, in its definition we exclude all diagrams 
with photons emitted and absorbed by the same electron. 
We split this contribution into the first-order and second-order parts,
\begin{align}
E_{\textrm{exch}}^{(7)} = \lbr H_{\rm{exch}}^{(7)}\rbr + E_\textrm{exch}^{\rm{sec}}\,,
\end{align}
where
\begin{align}
E_\textrm{exch}^{\rm{sec}} =&\ 2\,\bigg\langle H_\textrm{exch}^{(5)}\frac{1}{(E_0-H_0)'}H^{(4)}\bigg\rangle\,, \label{so-exch}
\end{align}
and
\begin{align}
H^{(5)}_{\rm exch} =&\  -\frac{7}{6\pi}\,\frac{1}{r^3} \,.
\end{align}
It is advantageous to express $\langle H_{\rm{exch}}^{(7)}\rangle$ using a set of operators $Q_i$
with $i=1\dots64$ which are suited for a numerical evaluation and are summarized in
Table~\ref{oprsQ1}. The first 50 of these operators were defined in
Refs.~\cite{pachucki:06:hesinglet, patkos:16:triplet, patkos:17:singlet}, whereas the remaining
14 operators are exclusive for the $\alpha^7\,m$ contribution.
The final expression for the photon-exchange contribution is
\begin{widetext}
\begin{eqnarray}
E^{(7)}_{\textrm{exch}} &=&
\frac{1}{\pi}\bigg\{\ln\frac{\alpha^{-2}}{2}\bigg( - \frac{22}{45}Z\,Q_3- \frac{19}{90} Q_{6T}-\frac{4}{15}Q_{10}
+ \frac{2}{15}Z\,Q_{18}
- \frac{1}{15}Z\,Q_{62}\bigg)
+\bigg(-\frac{772}{675} + \frac{22}{45}\ln2\bigg) Z\,Q_3
\nonumber\\
&&+\bigg(\frac{7937}{2700}-\frac{9}{10}\ln2\bigg)  Q_{6T}
+\bigg(-\frac{617}{1800}-\frac{8}{15}\ln2\bigg)Q_{10}+\bigg(\frac{841}{1800}+\frac23\ln2\bigg)Z\,  Q_{18}
+ \frac{31}{240} Q_{25}+ \frac{44}{45}Z\,Q_{52}
\nonumber\\
&&+\frac{8}{15} Q_{54} - \frac{33}{10} Q_{55}
- \frac{4}{15} Z\, Q_{58}   - \frac{7}{10} Q_{60} + \frac{7}{24} Q_{61}
 + \bigg(\frac{14}{225}+\frac{1}{15}\ln2\bigg)  Z\,Q_{62} + \frac{2}{15} Z\, Q_{63}\bigg\}
+E_\textrm{exch}^{\rm{sec}}\,.
\end{eqnarray}
The above expression was obtained by slightly simplifying our former result in
Ref.~\cite{patkos:20} with help of the following expectation value identity
\begin{align}\label{ident}
&\ p^i\,\biggl(\frac{\delta^{ij}}{r^3}-3\frac{r^i r^j}{r^5}\biggr)\,p^j
 =\frac{2\pi}{3}\,\vec{p}\,\delta^{3}(r)\,\vec{p}
 +\frac Z4\,\biggl(\frac{\vec{r}_1}{r_1^3}-\frac{\vec{r}_2}{r_2^3}\biggr)
\cdot\frac{\vec{r}}{r^3}-\frac{1}{2\,r^4}\,.
\end{align}
\end{widetext}
The expression (\ref{so-exch}) for $E_\textrm{exch}^{\rm{sec}}$ is finite but numerically
unstable. We thus regularize it as
\begin{eqnarray}
E_\textrm{exch}^{\rm sec} &=& 2\,\bigg\langle H_\textrm{exch}^{(5)}\frac{1}{(E_0-H_0)'}H_R\bigg\rangle
 \nonumber\\ &&
+ \frac{7}{6\pi}\Big[Q_9 ( Z\,Q_{53}- Q_7)
+  Q_{10} - Z\,Q_{59} \Big]\,,
\end{eqnarray}
where the regularized Breit operator $H_R$ is acting on ket-state $|\phi\rangle$ as
\begin{eqnarray}
H_R|\phi\rangle &=&\bigg(-\frac12(E_0-V)^2-\frac{Z}{4}\frac{\vec r_1\cdot\vec\nabla_1}{r_1^3}
-\frac{Z}{4}\frac{\vec r_2\cdot\vec\nabla_2}{r_2^3}\nonumber\\&&+\frac14\nabla_1^2\nabla_2^2
+\nabla_1^i\frac{1}{2r}\bigg(\delta^{ij}+\frac{r^i r^j}{r^2}\bigg)\,\nabla_2^j\bigg)|\phi\rangle\,, \label{13}
\end{eqnarray}
where $E_0$ is the nonrelativistic energy of $\phi$ and $V= - \frac{Z}{r_1} - \frac{Z}{r_2} +
\frac{1}{r}$. $H_R$ is equivalent to $H^{(4)}$ in the sense that their expectation values on
$\phi$ are the same.

\subsection{The radiative contribution}

The radiative correction $E^{(7)}_{\rm rad}$ consists of the one-loop self-energy, the one-loop  vacuum polarization,
the two-loop correction, and the three-loop correction,
\begin{equation}
E^{(7)}_{\rm rad} = E^{(7)}_{\rm SE} + E^{(7)}_{\rm VP} + E^{(7)}_{\rm rad2} + E^{(7)}_{\rm rad3}\,. \label{14}
\end{equation}
The one-loop self-energy contribution is
\begin{widetext}
\begin{eqnarray}
E_\textrm{SE}^{(7)} &=&\frac{1}{\pi}\bigg\{
-\frac{20}{9} E_0 E^{(4)} + \bigg(\frac{491}{1800}E_0 + \frac{3641}{3600} Z^2-\frac{1289}{360}Z^2\ln2 -\frac{5}{36} \pi^2Z^2
+\frac{8}{3}Z^2\ln^2 2+ \frac{5}{18} Q_7  + \frac54Z^2\,\zeta(3)\bigg) Z\, Q_1
\nonumber\\
&&- \frac{491+509Z}{1800}Z\, Q_3
+ \frac{509}{3600}Z\,Q_4  - \frac{1039}{1350} Q_{6T}
+\frac{10}{9} E^{(4)} Q_7
+ \bigg(\frac{403}{90}-\frac23\ln2\bigg)Q_{10}
 + \frac{10}{9} E_0 Z^2\,Q_{11}\nonumber\\
&&  + \frac{20}{9} E_0 Z^2 \,Q_{12} - \frac{20}{9} E_0 Z \,Q_{13}
-\frac{20}{9} Z^2 \,Q_{14}
+ \frac{20}{9} Z^3 \,Q_{15}- \frac{10}{9} Z^2 \,Q_{16} + \frac{10}{9} Z \, Q_{17}
- \frac{1271}{360} Z\, Q_{18}\nonumber\\
&&
+ \frac{5}{9} Z^2\, Q_{21}  + \frac59 Z^2\, Q_{22}
+ \frac{10}{9} Z\,Q_{24}+ \frac{1}{60} Q_{25} - \frac{5}{9}Z\,Q_{28}
+ \frac{779}{3600} Z\, Q_{51}+\frac{10}{9} E_0^2 Z \,Q_{53}-\frac43 Q_{54}\nonumber\\
&&
+ \frac{163}{120} Z^2\, Q_{57}
+ \ln\frac{\alpha^{-2}}{2}\bigg[-\frac83 E_0 E^{(4)}+\bigg(-\frac{E_0}{15} + \frac{53}{120}Z^2 + \frac53 Z^2\ln2
+\frac13 Q_7\bigg)Z\,Q_1 +\frac{1-11Z}{15}Z\,Q_3\nonumber\\
&& + \frac{11}{30}Z\,Q_4  - \frac{49}{45} Q_{6T}
+ \frac43 E^{(4)} Q_7  + 2\, Q_{10}
+\frac43 E_0 Z^2\,Q_{11} + \frac83 E_0 Z^2\, Q_{12} - \frac83 E_0 Z\,Q_{13}-\frac83 Z^2 \, Q_{14}\nonumber\\
&&
 + \frac83 Z^3 \, Q_{15} - \frac43 Z^2 \,Q_{16} + \frac43 Z\,Q_{17}
- \frac83 Z\,Q_{18} + \frac23 Z^2\,Q_{21} + \frac23 Z^2 \,Q_{22} + \frac43 Z\,Q_{24}- \frac23 Z\,Q_{28}\nonumber\\
&&
+ \frac{11}{30} Z\,Q_{51}+ \frac43 E_0^2 Z\,Q_{53}+Z^2 \,Q_{57}\bigg]
-\frac12\ln^2\frac{\alpha^{-2}}{2} Z^3\,Q_1\bigg\} + E_\textrm{SE}^{\rm sec}\,,
\end{eqnarray}
where $E^{(4)} = \big< H^{(4)}\big>$ is the Breit correction to the energy. The above formula for
$E_\textrm{SE}^{(7)}$ is obtained by simplifying our former result in Ref.~\cite{patkos:21} with
help of the identity (\ref{ident}). The second-order part $E_\textrm{SE}^{\rm sec}$ is
\begin{eqnarray}
E_\textrm{SE}^{\rm sec} &=& 2\bigg\langle H''^{(5)}\frac{1}{(E_0-H_0)'}H''^{(4)}\bigg\rangle
+\frac1\pi\bigg(\frac59+\frac23\ln\frac{\alpha^{-2}}{2}\bigg)\,
\bigg\langle H_R'\frac{1}{(E_0-H_0)'}H_R\bigg\rangle\,.
\end{eqnarray}
Here, the operators $H''^{(4)}$ and $H''^{(5)}$ are obtained, respectively, as the $\alpha^4$ and
$\alpha^5$ parts of the spin-dependent Breit Hamiltonian with anomalous magnetic moment (see,
e.g., Eq.~(1) of Ref.~\cite{pachucki:09:hefs}),
\begin{eqnarray}
H_{\rm fs} &=& H''^{(4)} + H''^{(5)} +O(\kappa^2)= H_B + H_C + H_D\,,\\
H_B & = & \bigg[
\frac{Z}{4}\biggl(\frac{\vec{ r}_1}{r_1^3}\times\vec{ p}_1+\frac{\vec{ r}_2}{r_2^3}\times\vec{ p}_2\biggr)\,(1+2\,\kappa)
-\frac{3}{4}\,\frac{\vec{ r}}{r^3}\times(\vec{ p}_1-\vec{ p}_2)\,\biggl(1+\frac{4\,\kappa}{3}\biggr)\bigg]\,\frac{\vec{\sigma}_1+\vec{\sigma}_2}{2}
 = (\vec Q_B + \kappa\,\vec Q'_B)\,\frac{\vec{\sigma}_1+\vec{\sigma}_2}{2}\,,\\
H_C& = & \bigg[
\frac{Z}{4}\biggl(\frac{\vec{ r}_1}{r_1^3}\times\vec{ p}_1-\frac{\vec{ r}_2}{r_2^3}\times\vec{ p}_2\biggr)\,(1+2\,\kappa)
+\frac{1}{4}\,\frac{\vec{ r}}{r^3}\times
(\vec{ p}_1+\vec{ p}_2)\bigg]\,\frac{\vec{\sigma}_1-\vec{\sigma}_2}{2}
= (\vec Q_C + \kappa\,\vec Q'_C)\,\frac{\vec{\sigma}_1-\vec{\sigma}_2}{2} \,,\label{50}\\
H_D  & = & \frac{1}{4}\left(
\frac{\vec{\sigma}_1\,\vec{\sigma}_2}{r^3}
-3\,\frac{\vec{\sigma}_1\cdot\vec{r}\,\vec{\sigma}_2\cdot\vec{r}}{r^5}\right)\,(1+\kappa)^2
=\big(Q^{ij}_D + \kappa\,Q'^{ij}_D\big)\,\frac{1}{2}\,\sigma_1^i\,\sigma_2^j + O(\kappa^2)
\,,
\end{eqnarray}
where $\kappa=\alpha/2\pi$ is anomalous magnetic moment correction. $H_R$ is defined in Eq. (\ref{13}), and $H_R'$ is
\begin{align}
H_R'|\phi\rangle = -2Z\bigg(\frac{\vec r_1\cdot\vec\nabla_1}{r_1^3}+\frac{\vec r_2\cdot\vec\nabla_2}{r_2^3}\bigg)|\phi\rangle\,.
\end{align}

Introducing the short-hand notations
\begin{align}
Q_A =&\  H_R\,,\\
Q'_A =&\ \bigg(\frac{5}{9}+\frac{2}{3}\,\ln\frac{\alpha^{-2}}{2}\bigg)\,H'_R\,,
\end{align}
we evaluate the second-order corrections as follows. After tracing out spins, we obtain for the
$2^3S_1$ state
\begin{eqnarray}
E(2^3S_1)^{\rm sec}_{\rm SE} &=& \frac{1}{\pi}\biggl\{ \langle 2^3S|Q'_A\,\frac{1}{(E_0-H_0)'}\,Q_A |2^3S\rangle
              +\frac{2}{3}\,\langle 2^3S|Q'^j_B\,\frac{1}{(E_0-H_0)'}\,Q_B^j
              |2^3S\rangle \nonumber \\
          &+& \frac{1}{3}\,\langle 2^3S|Q'^j_C\,\frac{1}{(E_0-H_0)'}\,Q_C^j |2^3S\rangle
              +\frac{1}{3}\,\langle 2^3S|Q'^{ij}_D\,\frac{1}{(E_0-H_0)'}\,Q_D^{ij} |2^3S\rangle\biggr\}\,.
\end{eqnarray}
A similar result holds for the $2^3P$ centroid,
\begin{eqnarray}
E(2^3P)^{\rm sec}_{\rm SE} &=& \frac{1}{\pi}\biggl\{\langle 2^3P^i| Q'_A\,\frac{1}{(E_0-H_0)'}\,Q_A |2^3P^i\rangle+
                     \frac{2}{3}\,\langle 2^3P^i| Q'^j_B\,\frac{1}{(E_0-H_0)'}\,Q_B^j
                     |2^3P^i\rangle \nonumber\\
                 &+& \frac{1}{3}\,\langle 2^3P^i| Q'^j_C\,\frac{1}{(E_0-H_0)'}\,Q_C^j |2^3P^i\rangle+
                 \frac{1}{3}\,\langle 2^3P^i| Q'^{jk}_D\,\frac{1}{(E_0-H_0)'}\,Q_D^{jk} |2^3P^i\rangle\biggr\}\,,
\end{eqnarray}
where we assumed the normalization $\langle 2^3P^i| 2^3P^i\rangle= 1$. This completes the
description of the first term in Eq.~(\ref{14}) which is the electron self-energy contribution.

The second term in Eq.~(\ref{14}) is  the one-loop vacuum polarization correction, for which we
obtained \cite{patkos:21}
\begin{eqnarray}
E_\textrm{VP}^{(7)} &=& \frac{1}{\pi}\bigg\{\frac{8}{15} E_0 E^{(4)}  + \bigg(-\frac{E_0}{105} + \frac{137}{1050}Z^2 - \frac{\pi^2}{54} Z^2 - \frac{1}{15} Q_7-\frac{1}{15}Z^2\,\ln\alpha^{-2}\bigg) Z\, Q_1
+\frac{1+13Z}{105}Z\, Q_3 - \frac{13}{210} Z\,Q_4\nonumber\\
&&+ \frac{13}{63} Q_{6T} - \frac{4}{15} E^{(4)} Q_7  - \frac{4}{15} E_0 Z^2\, Q_{11} - \frac{8}{15} E_0 Z^2\, Q_{12}
+\frac{8}{15} E_0 Z\, Q_{13} + \frac{8}{15} Z^2 \, Q_{14} - \frac{8}{15} Z^3 \, Q_{15}+ \frac{4}{15}Z^2 \, Q_{16}\nonumber\\
&& - \frac{4}{15}Z\,Q_{17}  - \frac{2}{15} Z^2\, Q_{21}
-\frac{2}{15} Z^2\, Q_{22} - \frac{4}{15} Z\, Q_{24} + \frac{2}{15} Z\, Q_{28}
 - \frac{13}{210} Z\, Q_{51} - \frac{4}{15} E_0^2 Z\, Q_{53}+ \frac{Z^2}{15} Q_{57}
\bigg\} + E_\textrm{VP}^{\rm sec}\,,\nonumber\\
\end{eqnarray}

\end{widetext}
with
\begin{equation}
E_\textrm{VP}^{\rm sec} = -\frac{2}{15\pi}\,
\bigg\langle H_R'\frac{1}{(E_0-H_0)'}H_R\bigg\rangle.
\end{equation}

Finally, the two-loop and three-loop radiative corrections are obtained from the known hydrogenic
results, keeping only the part proportional to the electron-nucleus contact interaction, whereas
the electron-electron contact interaction terms vanish because the nonrelativistic wave function
is antisymmetric with respect to the exchange $\vec r_1\leftrightarrow\vec r_2$. Therefore, the
two-loop correction is
\begin{align}\label{eq:QED2}
E^{(7)}_{\rm rad2} = \frac{Z^2}{2\pi^2}\,Q_1\, B_{50}\,,
\end{align}
where the coefficient $B_{50}$ is known only numerically, $B_{50} = -21.554\,47\,(13)$
\cite{yerokhin:18:hydr}. Similarly,  the three-loop radiative correction is given by
\cite{yerokhin:18:hydr}
\begin{align}\label{eq:QED3}
E^{(7)}_{\rm rad3} = &\ \frac{Z}{2\pi^3}\,Q_1\,
\bigg[ -{{568\,{a_4}}\over{9}}+{{85\,\zeta(5)}\over{24}}
-{{121\,\pi^{2}\,\zeta(3)}\over{72}} \nonumber\\
&\ -{{84\,071\,\zeta(3)}\over{2304}} -{{71\,\ln ^{4}2}\over{27}}
-{{239\,\pi^{2}\,\ln^{2}2}\over{135}}
 \nonumber\\&\hspace*{-5ex}
+{{4787\,\pi^{2}\,\ln 2}\over{108}}
+{{1591\,\pi^{4}}\over{3240}}
-{{252\,251\,\pi^{2}}\over{9720}}+{679\,441\over93\,312}
\bigg]\,,\nonumber\\
\end{align}
where $a_4 = \sum_{n=1}^\infty 1/(2^n\,n^4) = 0.517\,479\,061\dots$. This completes our
evaluation of the $\alpha^7\,m$ contribution.

\section{Estimation of $\bm{\alpha^8m}$ effects}

For the estimation of the radiative $\alpha^8m$ effects in helium, we employ the known hydrogenic
results and pretend that they are proportional to the electron-nucleus contact interaction.
Specifically, we use the results for the hydrogenic $2s$ state of He$^+$ \cite{yerokhin:18:hydr}
\begin{align} \label{eq:ma8a}
E^{(8+)}_{\rm rad1}({\rm hydr}) =&\ \frac{Z^7}{8\pi}\, \Big( 81.934_{\rm \,SE} + 1.890_{\rm\, VP}\Big)\,,
 \\
E^{(8+)}_{\rm rad2}({\rm hydr}) =&\ \frac{Z^6}{8\pi^2}\, \Big( -\frac8{27}\,\ln^3 [(Z\alpha)^{-2}]
 + 0.639\,\ln^2 [(Z\alpha)^{-2}] \nonumber\\ &\
+ 41.387\,\ln [(Z\alpha)^{-2}]  -81.1 \pm 10 \Big)\,,
 \label{eq:ma8b}
\end{align}
where the subscripts ``SE'' and ``VP'' denote the self-energy and vacuum-polarization
contributions, respectively. The three-loop contribution is small \cite{karshenboim:19} and thus
is neglected. The approximate $\alpha^8m$ corrections to the ionization energies of the $2^3S$
and $2^3P$ states of helium are obtained from the corresponding hydrogenic $2s$ contributions by
\begin{align} \label{eq:ma8c}
E^{(8+)} = E^{(8+)}({\rm hydr})\, \frac{\big< \delta^3(r_1) + \delta^3(r_2)\big> - \nicefrac{Z^3}{\pi}}{\nicefrac{Z^3}{8\pi}}\,.
\end{align}
Specifically, we get contributions of $0.158(52)$~MHz and $-0.048(16)$~MHz for the ionization
energies of the $2^3S$ and $2^3P$ states, correspondingly. We estimated the uncertainties to be
$1/3$ of the corresponding numerical values; this estimate can be improved further once the
$\alpha^7m$ contribution is verified.

\section{Finite nuclear size effect}
The last significant correction is due to the finite nuclear size, namely (in relativistic units)
\begin{align}\label{eq:34}
E_{\rm fns} =&\  \frac{2\,\pi}{3}\,Z\,\alpha\,\Big\langle\sum_a \delta^{(3)}(r_a)\Big\rangle\,R^2\,
\big[1-(Z\,\alpha)^2\,\ln(m\,R\,Z\,\alpha)\big]\,,
\end{align}
where $R$ is the root-mean-square nuclear charge radius, and the expectation value of the Dirac
$\delta$ functions is assumed to include the finite nuclear mass effects.

We note that Eq.~(\ref{eq:34}) includes relativistic effects in the form of the leading
logarithmic correction. Higher-order corrections to Eq.~(\ref{eq:34}) were investigated for
hydrogen-like atoms in Ref.~\cite{pachucki:18}. Crude scaling shows that for helium they are
negligible at the current level of precision and thus are neglected.

\section{Numerical method}
\begin{table*}
\caption{Second-order corrections for the $2\,^3\!S$ state; the prime on the sum means
exclusion of the reference state.\label{tab:sec:2S1}}
\begin{ruledtabular}
\begin{tabular}{lcdddd}
& \multicolumn{1}{c}{Intermediate}
   &   \multicolumn{1}{c}{$2\,^3\!S$}  \\
& \multicolumn{1}{c}{state}
   &   \multicolumn{1}{c}{}   \\ \hline\\[-5pt]
$\,\sum_n^{\prime}\frac1{E_0-E_n}{}\big<^3\!S \big|
 H_R' \big| n^3\!S\big>\, \big<n^3\!S\big| H_R
\big|^3\!S\big>$
            & $^3\!S$ &
             203.050\,945  \\[5pt]
$\,\sum_n^{\prime}\frac1{E_0-E_n}{}\big<^3\!S \big|
 H_\textrm{exch}^{(5)} \big| n^3\!S\big>\, \big<n^3\!S\big| H_R
\big|^3\!S\big>$
            & $^3\!S$ &
                          -0.030\,546  \\[10pt]
$\nicefrac2{3\pi}
\,\sum_n\frac1{E_0-E_n}{}\big<^3\!S \big|Q^{\prime
        i}_B\big|n^3\!P^i\big>\,\big< n^3\!P^j\big|Q_B^j \big| ^3\!S\big>$
                & $^3\!P^e$ &
                                                                          -0.003\,868 \\[5pt]
$\nicefrac1{3\pi}
 \,\sum_n\frac1{E_0-E_n}{}\big<^3\!S \big|Q^{\prime i}_C\big|n^1\!P^i\big>\,\big< n^1\!P^j\big|Q_C^j \big| ^3\!S\big>$
         & $^1\!P^e$ &
                                                                          -0.000\,195 \\[5pt]
$\nicefrac1{3\pi}
 \,\sum_n\frac1{E_0-E_n}{}\big<n^3\!D^{ij} \big|Q^{ij}_D\big|^3\!S\big>^2$
& $^3\!D^e$ &
                                                                          -0.001\,225 \\[5pt]
\end{tabular}
\end{ruledtabular}
\end{table*}
\begin{table*}
\caption{Second-order corrections for the $2\,^3\!P$  state (centroid). Normalization is according to $\langle P^i|P^i\rangle=
\langle D^{ij}|D^{ij}\rangle=\langle F^{ijk}|F^{ijk}\rangle=1$.\label{tab:sec:2P1}}
\begin{ruledtabular}
\begin{tabular}{lcddd}
& \multicolumn{1}{c}{Intermediate}
     &  \multicolumn{1}{c}{$2\,^3\!P$} & \\
& \multicolumn{1}{c}{state}
    &  \multicolumn{1}{c}{} & \\ \hline\\[-5pt]
$\,\sum_n^{\prime}\frac1{E_0-E_n}{}\big<^3\!P^i \big|
 H_R' \big| n^3\!P^i\big>\, \big<n^3\!P^k\big| H_R
\big|^3\!P^k\big>$
            & $^3\!P^o$ &
                                                                          190.798\,218\,(3) \\[5pt]
$\,\sum_n^{\prime}\frac1{E_0-E_n}{}\big<^3\!P^i \big|
 H_\textrm{exch}^{(5)} \big| n^3\!P^i\big>\, \big<n^3\!P^k\big| H_R
\big|^3\!P^k\big>$
            & $^3\!P^o$ &
                                                                           0.000\,059\,(2) \\[10pt]
$\nicefrac1{3\pi}
 \,\sum_n^{\prime}\frac1{E_0-E_n}{}\big<^3\!P^i \big|i\epsilon^{ijk}Q^{\prime j}_B\big|n^3\!P^k\big>
       \,\big< n^3\!P^l\big|i\epsilon^{lmn}Q_B^m \big| ^3\!P^n\big>$
            & $^3\!P^o$ &
                                                                          -0.008\,025\, \\[5pt]
$\nicefrac2{3\pi}
 \,\sum_n\frac1{E_0-E_n}{}\big<^3\!P^i \big|Q^{\prime j}_B\big|n^3\!D^{ij}\big>
       \,\big< n^3\!D^{lm}\big|Q_B^l \big| ^3\!P^{m}\big>$
           & $^3\!D^o$ &
                                                                          -0.000\,555\, \\[10pt]
$\nicefrac1{6\pi}
  \,\sum_n\frac1{E_0-E_n}{}\big<^3\!P^i \big|i\epsilon^{ijk}Q^{\prime
j}_C\big|n^1\!P^k\big>
       \,\big< n^1\!P^l\big|i\epsilon^{lmn}Q_C^m \big| ^3\!P^n\big>$
        & $^1\!P^o$ &
                                                                          -0.028\,515 \\[5pt]
$\nicefrac1{3\pi}
 \,\sum_n\frac1{E_0-E_n}{}\big<^3\!P^i \big|Q^{\prime j}_C\big|n^1\!D^{ij}\big>
       \,\big< n^1\!D^{lm}\big|Q_C^l \big| ^3\!P^{m}\big>$
                                                                       & $^1\!D^o$ &
                                                                          -0.000\,106 \\[10pt]
$\nicefrac1{5\pi}
 \,\sum_n^{\prime}\frac1{E_0-E_n}{}\big<n^3\!P^i \big|Q^{ik}_D\big|^3\!P^k\big>^2
       $
        & $^3\!P^o$ &
                                                                          -0.002\,429 \\[5pt]
$\nicefrac{2}{9\pi}
 \,\sum_n\frac1{E_0-E_n}{}\big<n^3\!D^{ij} \big|i\epsilon^{ikl} Q^{jk}_D\big|^3\!P^l\big>^2
       $
                                                                    & $^3\!D^o$ &
                                                                          -0.000\,015 \\[5pt]
$\nicefrac{1}{3\pi}
 \,\sum_n\frac1{E_0-E_n}{}\big<n^3\!F^{ijk} \big|Q^{ij}_D\big|^3\!P^{k}\big>^2
       $
                                                                    & $^3\!F^o$ &
                                                                          -0.000\,495 \\[5pt]
\end{tabular}
\end{ruledtabular}
\end{table*}

The spatial part of the helium wave function is expanded in a basis set of exponential functions
of the form \cite{korobov:00,korobov:02}
\begin{eqnarray}\label{wfunction}
\phi_i(r_1,r_2,r)&=&e^{-\alpha_i r_1-\beta_i r_2-\delta_i r}\pm(r_1\leftrightarrow r_2)\,,\\
\vec\phi_i(r_1,r_2,r)&=&\vec r_1\,e^{-\alpha_i r_1-\beta_i r_2-\delta_i r}\pm(r_1\leftrightarrow r_2)\,,
\end{eqnarray}
for the $S$ and $P$ states, correspondingly. The calculation of matrix elements of the
nonrelativistic Hamiltonian is performed with help of the formula
\begin{eqnarray}\label{eq:38}
\frac{1}{16\pi^2}\!\int d^3r_1\!\int d^3r_2\frac{e^{-\alpha r_1-\beta r_2-\delta r}}{r_1 r_2 r}=
\frac{1}{(\alpha+\beta)(\beta+\delta)(\delta+\alpha)}\,.\nonumber\\
\end{eqnarray}
The results for integrals with any additional powers of $r$ in the numerator can be obtained by
differentiation with respect to the corresponding parameter $\alpha$, $\beta$, or $\delta$.

Matrix elements of relativistic corrections involve integrals with additional inverse powers of
$r_1$, $r_2$, and $r$. Formulas for such integrals can be obtained by integrating
Eq.~(\ref{eq:38}) with respect to the corresponding nonlinear parameter. This leads to the
appearance of logarithmic and dilogarithmic functions; specifically,
\begin{widetext}
\begin{align}
\frac{1}{16\pi^2}\int d^3r_1\int d^3r_2\frac{e^{-\alpha r_1-\beta r_2-\delta r}}{r_1 r_2 r^2}=&\
\frac{1}{(\alpha+\beta)(\alpha-\beta)}\,\ln\bigg(\frac{\alpha+\delta}{\beta+\delta}\bigg) \,, \\
\frac{1}{16\pi^2}\int d^3r_1\int d^3r_2\frac{e^{-\alpha r_1-\beta r_2-\delta r}}{r_1^2 r_2 r^2}=&\
\frac{1}{2\,\beta}\bigg[\frac{\pi^2}{6}+\frac{1}{2}\,\ln^2\bigg(\frac{\alpha+\beta}{\beta+\delta}\bigg)
+{\rm Li}_2\bigg(1-\frac{\alpha+\delta}{\alpha+\beta}\bigg) + {\rm Li}_2\bigg(1-\frac{\alpha+\delta}{\beta+\delta}\bigg) \bigg]
\,.
\end{align}
Other integrals for relativistic corrections are obtained by differentiating the two basic
formulas above.

In our calculation of the $\alpha^7\,m$ contribution, we encounter operators involving $\ln
r+\gamma$, where $\gamma$ stands for the Euler's gamma constant. For the evaluation of these operators we
obtained the following formulas
\begin{align}
\frac{1}{16\,\pi^2}\,\int d^3r_1\,\int d^3 r_2\, e^{-\alpha\,r_1-\beta\,r_2-\delta\,r}\,4\,\pi\delta(r_1)\,\frac{(\ln r+\gamma)}{r} =&\
\frac{1-\ln(\beta+\delta)}{(\beta+\delta)^2}\,,
\end{align}
and
\begin{align}
\frac{1}{16\,\pi^2}\,\int d^3r_1\,\int d^3 r_2 \frac{e^{-\alpha\,r_1-\beta\,r_2-\delta\,r}}{r_1\,r_2\,r}\,(\ln r+\gamma) =&\
\frac{1}{(\alpha-\beta)\,(\alpha+\beta)}\biggl[\frac{\ln(\alpha+\delta)}{\alpha+\delta} - \frac{\ln(\beta+\delta)}{\beta+\delta}\biggr]\,,\\
\frac{1}{16\,\pi^2}\,\int d^3r_1\,\int d^3 r_2 \frac{e^{-\alpha\,r_1-\beta\,r_2-\delta\,r}}{r_1\,r_2\,r^2}\,(\ln r+\gamma) =&\
\frac{1}{2\,(\alpha-\beta)\,(\alpha+\beta)}\bigl[\ln^2(\beta+\delta) - \ln^2(\alpha+\delta)\bigr]\,,\\
\frac{1}{16\,\pi^2}\,\int d^3r_1\,\int d^3 r_2 \frac{e^{-\alpha\,r_1-\beta\,r_2-\delta\,r}}{r_1^2\,r_2\,r^2}\,(\ln r+\gamma) =&\
\frac{1}{2\,\beta}\bigg\{
\frac{1}{2}\,\ln\bigg(\frac{\alpha-\beta}{\alpha+\beta}\bigg)\,\big[\ln^2(\alpha+\delta) - \ln^2(\beta+\delta)\big] \nonumber \\ &\hspace{-40ex}
+ \ln(\alpha+\delta)\bigg[{\rm Li}_2\bigg(\frac{-\beta + \delta}{\alpha + \delta}\bigg) - {\rm Li}_2\bigg(\frac{\beta + \delta}{\alpha + \delta}\bigg)\biggr]
+ {\rm Li}_3\bigg(\frac{-\beta + \delta}{\alpha + \delta}\bigg) - {\rm Li}_3\bigg(\frac{\beta + \delta}{\alpha + \delta}\bigg)\bigg\}\,,
\end{align}
\end{widetext}
where the last formula is valid for $\alpha>\beta$. The result for $\alpha<\beta$ is obtained by
an analytic continuation with help of the identities
\begin{align}
{\rm Li}_2\big(-z\big) + {\rm Li}_2\big(-z^{-1}\big)  =&\  -\frac{\pi^2}{6}-\frac{\ln^2(z)}{2}\,, \\
{\rm Li}_3\big(-z\big) - {\rm Li}_3\big(-z^{-1}\big) = &\ -\frac{\pi^2}{6} \ln(z) -\frac{1}{6}\,\ln^3(z)\,.
\end{align}
In our calculation, we have derived explicit formulas for the expectation values of all $Q_i$
operators, and they involve the combination of the above expressions with the additional rational
function of $\alpha, \beta$, and $\delta$.

\section{Results}

Table~\ref{tab:1} presents our numerical results for the relativistic corrections to the Bethe
logarithm, obtained previously in Ref.~\cite{yerokhin:18:betherel}. Numerical values of the
second-order corrections are summarized in Table~\ref{tab:sec:2S1} for the $2^3S$ state and in
Table~\ref{tab:sec:2P1} for the $2^3P$ centroid. The uncertainties present for some of the matrix
elements are negligible at the level of uncalculated higher-order contributions. Expectation
values of various first-order operators are listed in Table~\ref{oprsQ1}. The matrix elements
$Q_i$ with $i\le 50$ have already been evaluated in our previous investigations (see Tables I and
II of Ref.~\cite{patkos:16:triplet}), whereas the operators with $i > 50$ are first encountered
in the present work. The numerical uncertainties for $Q_i$'s are smaller than the last digit
shown.

Table~\ref{tab:ma7} summarizes our calculation of the $\alpha^7m$ contributions to the energies
of the $2^3S$ and $2^3P$ states of helium. In order to obtain contributions to the ionization
energy, we need to subtract the corresponding corrections for the $1S$ state of the He$^+$ ion,
listed in the last column of the table. The hydrogenic formulas for $E^{(7)}_{\rm SE}({\rm
He}^+)$ and $E^{(7)}_{\rm L}({\rm He}^+)$ are obtained from
Refs.~\cite{pachucki:93,jentschura:05:sese} as follows
\begin{align}
E^{(7)}_{\rm SE}({\rm He}^+,1S)  = &\ \frac{Z^6}{\pi}\biggl\{
-\frac{121}{60}+\frac{5}{2}\, \zeta(3)-\frac{5}{18}\, \pi^2-\frac{61}{90}\, \ln2\nonumber \\ &\
-3\, \ln^2 2 +\ln(Z\,\alpha)\, \bigg[\frac{163}{30}-4\,\ln(2\,\Lambda)\bigg]\nonumber \\ &\
-\frac{5}{3}\,\ln\Lambda-\frac{22}{3}\, \ln 2\, \ln\Lambda+\ln^2\Lambda\biggr\}\,,\\
E^{(7)}_{\rm L}({\rm He}^+,1S) =&\  \frac{Z^6}{\pi}\biggl\{
\beta + \biggl(\frac{5}{3} + \frac{22}{3}\,\ln2\biggr) \ln\biggl[\frac{\Lambda}{(Z\,\alpha)^2}\biggr] \nonumber \\ &\ - \ln^2\biggl[\frac{\Lambda}{(Z\,\alpha)^2}\biggr] \biggr\}\,,
\end{align}
where $\beta = \beta_1+\beta_2+\beta_3 = 27.25990948$ and $Z=2$. The sum $E^{(7)}_{\rm SE}({\rm
He}^+) + E^{(7)}_{\rm L}({\rm He}^+)$ does not depend on the cutoff parameter $\Lambda$. In order
to be consistent with our present calculations for atomic He, one should set the cutoff parameter
as $\Lambda\rightarrow\alpha^2$.

We note a strong cancellation between the He and He$^+$ corrections, which reflects the fact that
the dominant contribution to the $2^3\!S$ and $2^3\!P$ energies comes from the $1s$ electron. The
resulting $\alpha^7m$ correction to the ionization energy is in agreement with our previous
approximate predictions \cite{pachucki:17:heSummary} based on the known He$^+$ Lamb shift.

Table~\ref{tab:5} summarizes all known theoretical contributions to the ionization energies of
the $2^3\!S$ and $2^3\!P$ states of helium. The contributions up to order $\alpha^6\,m$
correspond to those from our review \cite{pachucki:17:heSummary}, with the updated value of the
Rydberg constant \cite{nist:web}. The finite nuclear size correction is calculated with the
charge radius obtained from the recent measurement of the muonic helium Lamb shift
\cite{krauth:21}. We find that the effects of order $\alpha^7\,m$ and $\alpha^8\,m$ shift the
$2^3S-2^3P$ transition frequency by $-8.447$~MHz and $0.206\,(54)$~MHz, respectively.

Table~\ref{tab:transition} compares our final theoretical predictions with experimental results.
There are three accurately measured transitions in He that involve the $2^3S$ and $2^3P$ states.
The theoretical transition energy $E(2^3S-2^3P)_{\rm theo}=276\,736\,495.620\,(54)$~MHz is in
very good agreement with the experimental result $E(2^3S-2^3P)_{\rm
exp}=276\,736\,495.600\,0\,(14)$~MHz from Ref.~\cite{zheng:17}, while for the other two
transitions, $2^3S-3^3D_1$ and $2^3P_0-3^3D_1$, theory and experiment disagree by about $0.5$
MHz.

\section{Discussion}

The theoretical energies contain the nuclear charge radius $R$ as a parameter, through the finite
nuclear size correction given by Eq.~(\ref{eq:34}). By comparing the theoretical predictions with
high-precision experimental results (particularly, the $2^3S-2^3P$ transition energy
\cite{zheng:17}), one can determine $R$. The present theoretical accuracy is in principle
sufficient for a determination of the nuclear radius with an accuracy of about 1\%. However, the
unexplained discrepancy between theory and experiment for the $2^3S-3^3D$ and $2^3P-3^3D$
transitions does not allow us to do this.

Disagreements between theory and experiment for transitions involving $D$ states have already
been reported previously \cite{pachucki:17:heSummary,wienczek:19,yerokhin:20:dstates}. The
present calculation reduces the discrepancy for triplet states from 1~MHz to 0.5~MHz. However,
the theoretical uncertainty due to uncalculated higher-order effects is now reduced by an order
of magnitude, so the relative discrepancy with experiment increased drastically, reaching
$15\,\sigma$ for the $2^3P_0-3^3D_1$ transition.

Bearing in mind the two different measurements, both of which show similar deviations from
theory, we conclude that the most plausible explanation of the discrepancy would be some unknown
theoretical contribution shifting the $2^3S$ and $2^3P$ states by approximately the same value.
For this reason we postpone the determination of the $\alpha$-particle charge radius by means of the
atomic spectroscopy until this unknown correction or a mistake in our calculations is identified.

\begin{acknowledgments}
K.P. and V.P.  acknowledge support from the National Science Center (Poland) Grant No. 2017/27/B/ST2/02459.
V.A.Y. acknowledges support from the Russian Science Foundation (Grant No. 20-62-46006).
V.P. acknowledges additional support from the Czech Science Foundation - GA\v{C}R (Grant No. P209/18-00918S)
\end{acknowledgments}

\begin{table*}
\renewcommand{\arraystretch}{0.92}
\caption{Expectation values of various operators for the $2^3\!S$ and $2^3\!P$ states,
$\vec P = \vec{p}_1+\vec{p}_2$, $\vec p = \nicefrac12
\big(\vec{p}_1-\vec{p}_2\big)$.}
\label{oprsQ1}
\begin{ruledtabular}
\begin{tabular}{lldddd}
 &  &   2^3S  &  2^3P & \\ \hline
$Q_1 $ & $4 \pi \delta^3 (r_1)$   				     &  16.592\,071        & 15.819\,309  \\
$Q_2 $ & $4 \pi \delta^3 (r)$               	             &   0 	     	 &  0  \\	
$Q_3 $ & $4 \pi \delta^3(r_1)/r_2$                  	     &   4.648\,724    &  4.349\,766 \\	
$Q_4 $ & $4 \pi \delta^3(r_1)\, p_2^2$ 	                     &   2.095\,714    &  4.792\,830 \\
$Q_5 $ & $4 \pi \delta^3(r)/r_1$				     &   0	           &  0 \\
$Q_{6T} $ & $4 \pi\,\vec{p}\,\delta^3(r)\,\vec{p} $		     &   0.028\,099    &  0.077\,524\\
$Q_7 $ & $1/r$						     &   0.268\,198           &  0.266\,641 \\
$Q_8 $ & $1/r^2$						     &   0.088\,906       &  0.094\,057 \\
$Q_9 $ & $1/r^3$                    	                     &   0.038\,861    &  0.047\,927 \\
$Q_{10}$ & $1/r^4$                  	                     &   0.026\,567    &  0.043\,348 \\
$Q_{11}$ & $1/r_1^2$                	                     &   4.170\,446    &  4.014\,865 \\
$Q_{12}$ & $1/(r_1 r_2)$            	                     &   0.560\,730    &  0.550\,342  \\
$Q_{13}$ & $1/(r_1 r)$              	                     &   0.322\,696    &  0.317\,639 \\
$Q_{14}$ & $1/(r_1 r_2 r)$          	                     &   0.186\,586    &  0.198\,346 \\
$Q_{15}$ & $1/(r_1^2 r_2)$					     &   1.242\,704    &  1.196\,631 \\
$Q_{16}$ & $1/(r_1^2 r)$					     &   1.164\,599    &  1.109\,463 \\
$Q_{17}$ & $1/(r_1 r^2)$   					     &   0.112\,360    &  0.121\,112 \\
$Q_{18}$ & $(\vec{r}_1\cdot\vec r)/(r_1^3 r^3)$                  &   0.011\,331    &  0.030\,284 \\
$Q_{19}$ & $(\vec{r}_1\cdot\vec r)/(r_1^3 r^2)$                  &   0.054\,635    &  0.075\,373 \\
$Q_{20}$ & $r_1^i r_2^j(r^i r^j-3\delta^{ij}r^2)/(r_1^3 r_2^3 r)$&   0.027\,082    &  0.090\,381 \\
$Q_{21}$ & $p_2^2/r_1^2$					     &   0.751\,913    &  1.410\,228 \\
$Q_{22}$ & $\vec{p}_1/r_1^2\, \vec{p}_1$			     &  16.720\,479    & 15.925\,672 \\
$Q_{23}$ & $\vec{p}_1/r^2\, \vec{p}_1$			     &   0.243\,754    &  0.279\,229 \\
$Q_{24}$ & $p_1^i\,(r^i r^j+\delta^{ij} r^2)/(r_1 r^3)\, p_2^j$  &   0.002\,750    & -0.097\,364 \\
$Q_{25}$ & $P^i\, (3 r^i r^j-\delta^{ij} r^2)/r^5\, P^j$	     &   0.062\,031    & -0.060\,473 \\
$Q_{26}$ & $p_2^k \,r_1^i\,/r_1^3 (\delta^{jk} r^i/r - \delta^{ik} r^j/r -\delta^{ij} r^k/r -r^i
r^j
r^k/r^3)\, p_2^j$		     &  -0.009\,102    &  0.071\,600\\
$Q_{27}$ & $p_1^2\, p_2^2$					     &   0.488\,198    &  1.198\,492 \\
$Q_{28}$ & $p_1^2\,/r_1\, p_2^2$				     &   1.597\,727    &  3.883\,405 \\
$Q_{29}$ & $\vec{p}_1\times\vec{p}_2\,/r\,\vec{p}_1\times\vec{p}_2$
							     &   0.070\,535    &  0.399\,306 \\
$Q_{30}$ & $p_1^k \,p_2^l\,(-\delta^{jl} r^i r^k/r^3 - \delta^{ik} r^j r^l/r^3 +3r^i r^j r^k
r^l/r^5)\,
p_1^i\, p_2^j$			     &  -0.034\,780    & -0.187\,305 \\
$Q_{31}$ & $4 \pi \delta^3(r_1)\, \vec{p}_1\cdot\vec{p}_2$       &   0.040\,294 & -0.457\,224   \\
$Q_{32}$ & $(\vec{r}_1\cdot\vec{r}_2)/(r_1^3 r_2^3)$             &  -0.005\,797 & -0.032\,383   \\
$Q_{33}$ & $\vec{p}_1\cdot\vec{p}_2$			     &   0.007\,442 & -0.064\,572  \\
$Q_{34}$ & $\vec{P}\,/r_1\,\vec{P}$				     &   4.974\,707 &  4.730\,359  \\
$Q_{35}$ & $\vec{P}\,/r\,\vec{P}$				     &   1.232\,372 &  1.127\,146   \\
$Q_{36}$ & $\vec{P}\,/r_1^2\,\vec{P}$                            &  17.504\,835 & 16.972\,775   \\
$Q_{37}$ & $\vec{P}\,/(r_1 r_2)\,\vec{P}$			     &   2.489\,592 &  2.291\,176   \\
$Q_{38}$ & $\vec{P}\,/(r_1 r)\,\vec{P}$			     &   1.454\,007 &  1.350\,214   \\
$Q_{39}$ & $\vec{P}\,/r^2\,\vec{P}$				     &   0.438\,804 &  0.413\,144   \\
$Q_{40}$ & $p_1^2\,p_2^2\,P^2$				     &  10.324\,509 & 24.527\,699   \\
$Q_{41}$ & $P^2\,p_1^i\, (r^i r^j + \delta^{ij} r^2)/r^3 \, p_2^j$
							     &   0.151\,748 &  0.067\,201   \\
$Q_{42}$ & $p_1^i\,(r_1^i r_1^j + \delta^{ij} r_1^2)/r_1^4 \,  P^j$
						             & 33.461\,709  & 31.489\,835  \\
$Q_{43}$ & $p_1^i\,(r_1^i r_1^j + \delta^{ij} r_1^2)/(r_1^3 r_2)\,  P^j$
						             &  2.486\,269  & 2.217\,310  \\
$Q_{44}$ & $p_1^i\,p_2^k\,(r_1^ir_1^j+\delta^{ij}r_1^2)/r_1^3\,p_2^k\, P^j$
							     &  1.100\,915  & 2.527\,505   \\
$Q_{45}$ & $p_2^i(r^i r^j+\delta^{ij} r^2)(r_1^jr_1^k
+\delta^{jk} r_1^2)/(r_1^3 r^3)\, P^k$	 	             &  0.540\,877  & 0.467\,623  \\
$Q_{46}$ & $p_1^i(r_1^i r_1^j+\delta^{ij} r_1^2)(r_2^jr_2^k
+\delta^{jk} r_2^2)/(r_1^3 r_2^3)\, p_2^k$		     &   0.006\,782 & -0.201\,826   \\
$Q_{47}$ & $(\vec{r}_1\cdot\vec{r}_2)/(r_1^3 r_2^2)$             &  -0.008\,117 & -0.028\,621   \\
$Q_{48}$ & $r_1^i r^j(r_1^i r_1^j-3\delta^{ij} r_1^2)/(r_1^4r^3)$&  -0.036\,861 & -0.057\,404   \\
$Q_{49}$ & $r_1^ir^j(r_2^ir_2^j-3\delta^{ij}r_2^2)/(r_1^3r_2r^3)$&  -0.089\,086 & -0.126\,780   \\
$Q_{50}$ & $p_2^k\,r_1^i/r_1^3\,(\delta^{jk}r_2^i/r_2-\delta^{ik}r_2^j/r_2
-\delta^{ij}r_2^k/r_2-r_2^ir_2^jr_2^k/r_2^3)\,p_2^j$	     &   0.005\,856 & -0.092\,036   \\
$Q_{51} $ & $4 \pi\,\vec p_1\,\delta^3(r_1)\,\vec p_1 $	         &  0.009\,993  &  0.270\,964 \\
$Q_{52} $ & $4 \pi \delta^3(r_1)/r_2\,(\ln r_2+\gamma)$          &  8.125\,982  &  7.514\,290\\	
$Q_{53} $ & $1/r_1$						                         &  1.154\,664  &  1.133\,242 \\
$Q_{54}$ & $1/r^4 (\ln r+\gamma)$                  	             &  0.015\,481  &  0.009\,473\\
$Q_{55}$ & $1/r^5$                  	                         &  0.017\,580  &  0.027\,240 \\
$Q_{56}$ & $1/r_1^3$   					                         &-23.022\,535  &-21.886\,142 \\
$Q_{57}$ & $1/r_1^4$   					                         & 25.511\,837  & 24.525\,751 \\
$Q_{58}$ & $(\vec{r}_1\cdot\vec r)/(r_1^3 r^3)(\ln r+\gamma)$    &  0.026\,515  &  0.038\,795\\
$Q_{59}$ & $1/(r_1 r^3)$				                         &  0.051\,914  &  0.069\,729 \\
$Q_{60}$ & $\vec p\,/r^3\, \vec p$				                 &  0.072\,885  &  0.093\,877 \\
$Q_{61}$ & $\vec P\,/r^3\, \vec P$				                 &  0.211\,990  &  0.226\,284 \\
$Q_{62}$ & $r^i r^j(\delta^{ij}r_1^2-3r_1^i r_1^j)/(r_1^5 r^3)$	 & -0.017\,688  & -0.051\,696 \\
$Q_{63}$ & $r^i r^j(\delta^{ij}r_1^2-3r_1^i r_1^j)/(r_1^5 r^3) (\ln r+\gamma)$				
                                                                 & -0.045\,609  & -0.045\,395\\
$Q_{64}$ & $p^i (\delta^{ij}r^2-3r^i r^j)/r^5 p^j$               &  0.002\,731  &  0.021\,530 \\
\end{tabular}
\end{ruledtabular}
\end{table*}

\begin{table}
\caption{Numerical results for individual contributions to $E^{(7)}$ for the $2^3\!S$ and
$2^3\!P$ (centroid) states of helium, in units of $\alpha^7m$ if not specified explicitly. \label{tab:ma7}
}
\begin{ruledtabular}
\begin{tabular}{l...}
\multicolumn{1}{l}{Term} & \multicolumn{1}{c}{$2^3S$} & \multicolumn{1}{c}{$2^3P$} & \multicolumn{1}{c}{He$^+(1S)$}\\
\hline\\[-5pt]
$E_L^{(7)}$         			          &  -804.x306\,(5)  &  -767.x828\,(10) &  -785.x107\\
$E^{(7)}_{\rm SE}$        				  &  -379.x061       &  -359.x257       &  -367.x554\\
$E^{(7)}_{\rm VP}$						  &   -36.x094       &   -34.x381       &   -34.x716\\
$E^{(7)}_{\rm rad2}$				      &   -72.x471       &   -69.x096       &   -69.x885\\
$E^{(7)}_{\rm rad3}$  				      &     0.x223       &     0.x213       &     0.x215\\
$E^{(7)}_{\rm exch}$   				      &   -10.x639       &    -9.x950       &     0.x000\\
$E^{(7)}$   				  			  & -1302.x348\,(5)  & -1240.x301\,(10) & -1257.x046\\
$E^{(7)}$\,[MHz]						  &  -177.x320\,(1)  &  -168.x872\,(1)  &  -171.x152\\
${\rm He} - {\rm He}^+$\,[MHz] 		      &    -6.x168\,(1)   &    2.x280\,(1)       \\
Prev. theory \cite{pachucki:17:heSummary}
																		&  -5.x2\,(1.3)  & 2.x9\,(0.7)
\end{tabular}
\end{ruledtabular}
\end{table}

\begin{table*}
\caption{Breakdown of theoretical contributions to the ionization (centroid) energies of the $2^3S$ and $2^3P$ states of $^4$He, in MHz.
$R_{\infty}c = 3.289\,841\,960\,250\,8(64) \times 10^{15}$~Hz \cite{nist:web}, $M/m_e = 7294.299\,541\,42\,(24)$ \cite{nist:web},
$1/\alpha = 137.035\,999\,206\,(11)$ \cite{morel:20}, $R = 1.678\,24\,(83)$~fm \cite{krauth:21}.
NS denotes the finite nuclear size correction; NP stands for the nuclear polarizability correction. 
The uncertainty of the theoretical $\alpha^2$ contribution comes from the Rydberg constant; the uncertainty of the
finite nuclear size correction comes from the nuclear radius.
\label{tab:5}}
\begin{ruledtabular}
\begin{tabular}{l w{13.6}w{10.6}w{6.6}w{6.6}w{10.6}}
 \multicolumn{1}{c}{$ $}
 & \multicolumn{1}{c}{$(m/M)^0$}
         & \multicolumn{1}{c}{$(m/M)^1$}
              & \multicolumn{1}{c}{$(m/M)^2$}
                  & \multicolumn{1}{c}{$(m/M)^3$}
                        & \multicolumn{1}{c}{Sum}\\
\hline\\[-5pt]
 ${ 2^3S:}$      \\[1pt]
$\alpha^2$ &        -1\,152\,953\,922.384\,(2)  & 164\,775.354        & -30.620             & 0.006             & -1\,152\,789\,177.644\,(2)     \\
$\alpha^4$ &                 -57\,629.312       &        4.284        &  -0.001             &                   &          -57\,625.029          \\
$\alpha^5$ &                   3\,999.431       &       -0.800        &                     &                   &            3\,998.632          \\
$\alpha^6$ &                       65.235       &       -0.030        &                     &                   &                65.205            \\
$\alpha^7$ &                       -6.168\,(1)  &                     &                     &               	&                -6.168\,(1)     \\
$\alpha^8$ &                        0.158\,(52) &                     &                     &             		&                 0.158\,(52)     \\
NS         &                        2.616\,(3)  &                     &                    &                    &                 2.616\,(3)        \\
NP         &                       -0.001       &                     &                    &                    &                -0.001        \\
Total      &                                    &                     &                    &                    & -1\,152\,842\,742.231\,(52)   \\
Theory 2017 \cite{pachucki:17:heSummary} &      &                     &                    &                    & -1\,152\,842\,741.4\,(1.3)   \\[5pt]
 ${ 2^3P:}$      \\[1pt]
$\alpha^2$ &        -876\,178\,284.857\,(2)    & 61\,871.895         & -25.840            & 0.006              & -876\,116\,438.795\,(2)      \\
$\alpha^4$ &               11\,436.878         &      11.053         &   0.002            &                    &        11\,447.932         \\
$\alpha^5$ &               -1\,234.732         &      -0.614         &                    &                    &        -1\,235.346           \\
$\alpha^6$ &                   -21.833         &      -0.001         &                    &                    &            -21.835             \\
$\alpha^7$ &                     2.280\,(1)    &                     &                    &                    &              2.280\,(1)    \\
$\alpha^8$ &                    -0.048\,(16)   &                     &                    &                    &             -0.048\,(16)    \\
NS         &                    -0.799\,(1)    &                     &                    &                    &             -0.799\,(1)         \\
NP         &                     0.000         &                     &                    &                    &              0.000 \\
Total      &                                   &                     &                    &                    & -876\,106\,246.611\,(16)    \\
Theory 2017 \cite{pachucki:17:heSummary}  &    &                     &                    &                    & -876\,106\,246.0\,(7)    \\
\end{tabular}
\end{ruledtabular}
\end{table*}

\begin{table*}
\caption{Comparison of experimental results for various transitions with theoretical predictions, in MHz. \label{tab:transition}}
\begin{center}
\begin{ruledtabular}
\begin{tabular}{l ll l l}
\centt{Transition}
  & \centt{Theory} & \multicolumn{2}{c}{Experiment} & \multicolumn{1}{c}{Difference} \\ \hline\\[-7pt]
$2^3S$--$3^3D_1\,$ & 786\,823\,849.540\,(52)$^a$ & 786\,823\,850.002\,(56) \, &\cite{dorrer:97}      & $-$0.462\,(76)\\
$2^3P_0$--$3^3D_1\,$ & 510\,059\,754.863\,(16)$^{a,b}$ & 510\,059\,755.352\,(28) \, &\cite{luo:16}   & $-$0.489\,(32)\\
$2^3P$--$2^3S\,$ & 276\,736\,495.620\,(54) & 276\,736\,495.600\,0\,(14) \, &\cite{zheng:17}$^b$      & $\ $   0.020\,(54)\\
\end{tabular}
\end{ruledtabular}
\end{center}

$^a$ using theoretical energy  $E(3^3D_1) = 366\,018\,892.691\,(23)$ from Ref.~\cite{yerokhin:20:dstates},\\
$^b$ using theoretical results for the $2^3P$ fine structure from Ref.~\cite{pachucki:11}.
\end{table*}


\end{document}